\documentclass[a4paper,fleqn,usenatbib]{mnras}

\usepackage{newtxtext,newtxmath}
\usepackage[T1]{fontenc}
\usepackage{ae,aecompl}

\usepackage{graphicx}	% Including figure files
\usepackage{amsmath}	% Advanced maths commands
\usepackage{amssymb}	% Extra maths symbols

\title[Dispersion of the solar magnetic flux]{Dispersion of the solar magnetic flux in undisturbed photosphere as derived from SDO/HMI data}

\author[V.I.Abramenko]{
Valentina I. Abramenko\thanks{E-mail: vabramenko@gmail.com (VIA)}
\\
Crimean Astrophysical Observatory, Russian Academy of Science, Nauchny, Bakhchisaray,  298409, Crimea, Russia
}

\date{Accepted XXX. Received YYY; in original form ZZZ}

\pubyear{2015}

\begin{document}
\label{firstpage}
\pagerange{\pageref{firstpage}--\pageref{lastpage}}
\maketitle

\begin{abstract}
To explore the magnetic flux dispersion in the undisturbed solar photosphere, magnetograms acquired by Helioseismic and Magnetic Imager (HMI) onboard the Solar Dynamic Observatory (SDO)  
were utilized. Two areas, a coronal hole area (CH) and an area of super-granulation pattern, SG, were analyzed. We explored the displacement and separation spectra and the behavior of the turbulent diffusion coefficient, $K$. The displacement and separation spectra are very similar to each other.
Small magnetic elements (of size  3-100 squared pixels and the detection threshold of 20 Mx sm$^{-2}$)  in both CH and SG areas disperse in the same way  and they are more mobile than the large elements (of size 20-400 squared pixels and the detection threshold of 130 Mx sm$^{-2}$).  The regime of super-diffusivity is found for small elements ($\gamma  \approx 1.3 $ and $K$ growing from $\sim$100 to $\sim$ 300 km$^2$ s$^{-1}$). 
Large elements in the CH area are scanty and show super-diffusion with $\gamma \approx 1.2$ and $K$ = (62-96) km$^2$ s$^{-1}$ on rather narrow range of 500-2200 km. Large elements in the SG area demonstrate two ranges of linearity and two diffusivity regimes: sub-diffusivity on scales (900-2500) km with $\gamma=0.88$ and $K$ decreasing from $\sim$130 to $\sim$100 km$^2$ s$^{-1}$, and super-diffusivity on scales (2500-4800) km with $\gamma \approx 1.3$ and $K$ growing from $\sim$140 to $\sim$200 km$^2$ s$^{-1}$. 
Comparison of our results with the previously published shows that there is a tendency of saturation of the diffusion coefficient on large scales, i.e., the turbulent regime of super-diffusivity is gradually replaced by normal diffusion.

\end{abstract}


%%%%%%%%%%%%%%%%%%%%%%%%%%%%%%%%%%%%%%%%%%%%%%%%%%

%%%%%%%%%%%%%%%%% BODY OF PAPER %%%%%%%%%%%%%%%%%%

\section{Introduction}

Solar magnetic fields are the key signature of solar activity, the major contributor to the near-Earth space weather and they as well control the total solar irradiance (TSI). A closer look at the solar photosphere reveals chaos of continuously renewing mixed-polarity magnetic elements spanning all spatial scales down to the resolution limits of modern instrumentation.
Flux transport models strive to describe the ceaselessly changing distribution of the magnetic flux driven by convection, differential rotation and meridional flows. Particular attention is paid to the dispersal of the magnetic flux. For example, how does the eroded magnetic field of a sunspot or a pore move through the magnetic network from the near solar equator to the poles? What are parameters of this process? Magnetic flux dispersal is usually described in terms of turbulent magnetic diffusivity. The corresponding diffusion coefficient characterizes mobility of magnetic elements and is a free parameter in the existing models of solar dynamo and magnetic flux transport, and it eventually affects reconstructed values of TSI.

Turbulent magnetic diffusivity - key parameter of the models - is the most poorly constrained parameter theoretically and observationally. The competition between diffusion and advection processes determines the solar cycle memory and thus affects the prediction of an oncoming cycle \citep[][]{Yeates2008}.
The flux transport models usually adopt a constant value of the diffusion coefficient, which substantially differs from one model to another: from 5 km$^2$ s$^-1$ \citep[][]{Jouve_Brun2007} to 600 km$^2$ s$^-1$  \citep[][]{Jiang2010,Wang2002}. Solar observations produce values below 300-350 km$^2$ s$^-1$ \citep[][]{Schrijver1996,Utz2010, Manso2011}, which strongly depend on the characteristic spatial and temporal scales of the utilized data. Accurate measurements of the diffusion coefficient are critical for further progress. Understanding of the diffusivity scale dependence is needed to calibrate the diffusivity profiles in theoretical dynamo models \citep[][]{Dikpati_Char1999, Rempel2006, Jouve2008, Pipin_Kos2011}.

Hinode data allowed to detect ubiquitous transverse and fine mixed polarity fields \citep[e.g.,][]{Lites2008,deWijn2008,deWijn2009,Ishikawa2010,Ishikawa_Tsu2010}, which led to an idea that an additional mechanism of magnetic field generation, such as local turbulent dynamos \citep[e.g.,][]{Petrovay2001} should be at work. It is a tremendously difficult task to observationally prove the existence of the local turbulent dynamo \citep{Stenflo2012} and it was only indirectly argued \citep[e.g.,][]{Lites2011, Abramenko2012}. MHD-simulations of solar magnetoconvection can help here because of the possibility to monitor the accumulated magnetic energy and to estimate the efficiency of local dynamo effects. So far they provide an affirmative answer - the local turbulent dynamo seems to work \citep[e.g.,][]{Stein_Nor2000, Boldyrev_Cat2004, Vogler_Sch2007, Pietarila2009}. In these models the diffusion is not input directly. However, it can be inferred from simulations and then compared to observations, which is vitally important for further progress in the field.

Granular and supergranular turbulent magnetic diffusion is also considered to be a constant in the flux transport equations \citep[][]{Wang2005}. Integration of the equations over many cycles allows to derive the time dependence of the total photospheric flux. The total flux, in turn, largely determines the TSI and is thus needed to reconstruct the solar radiative input to the Earth climate, as well as for future predictions \citep[][]{Foukal2012}.

The commonly accepted mechanism for transporting the magnetic flux over the
solar surface on small scales is random walk, or, normal diffusion,
when the mean-squared displacement of flow tracers varies with time, $\tau$, as
$\langle (\Delta l)^2 \rangle = 4K\tau \sim \tau^{\gamma}$, where $K$ is the
diffusion coefficient (a scalar), and $\gamma=1$, e.g., \citet[][]{Monin1975}. Generally, when index $\gamma$
deviates from unity, diffusion is called anomalous diffusion. More
specifically, a regime with $\gamma>1$ is called super-diffusive,
while $\gamma<1$, indicates sub-diffusive.
Parameters $\langle (\Delta l)^2 \rangle $ and $\gamma$, generally
derived from observations, allow us to determine the diffusion coefficient as a
function of time and spatial scales \citep[][]{Abramenko2011}.

Before the Hinode-era, it was acknowledged that the diffusion coefficient can sometimes vary depending on the data quality and scale of interest (see a review by \citet{Berger1998}). Recent researches based on the new-generation of solar instrumentation \citep{Abramenko2011, Gianna2013, Gianna2014a, Gianna2014b, Keys2014, Yang2015, Jafar2017}, strongly suggests that the diffusion coefficient is not constant and varies in direct proportion to the spatial and time scales suggesting the turbulent regime of super-diffusivity in the photosphere. 

Magnetic bright points are frequently used as tracers of the magnetic flux \citep[][]{Keys2014, Yang2015, Jafar2017}, and only in the series of publications by \citet{Gianna2013,Gianna2014a, Gianna2014b} dispersal of magnetic flux elements were considered. A considerable scattering in magnitudes of the observed index $\gamma$ can be stated. Thus, for the weakest magnetic environment of quiet sun, \citet{Jafar2017} on the basis of 121 tracers report $\gamma =1.9 \pm 0.7$ for the time scale interval of approximately (200-1000) s; \citet{Keys2014} from 851 tracers found $\gamma = 1.2 \pm 0.2$ for the very short time intervals of approximately 4-100 s; \citet{Yang2015} argue that index $\gamma$, as measured on scales below 300 s, decreases from 1.7 to 1.3 as the magnetic field increases from 100 to 450 G.  \citet{Gianna2014a} analyzing displacement spectra of magnetic elements in the interior of a supergranula found $\gamma = 1.44$ on scales of approximately 100-10000 s (in the spatial domain, this corresponds to approximately 150-4000 km, see Fig 3 in \citet{Gianna2014a}. For supergranula boundary elements, they found a break on scales of about 600 km with $\gamma = 1.27$ below and 1.08 above. Authors suggest that the lower diffusivity in the network areas facilitate the amplification of the magnetic field therein. 

In \citet{Gianna2014b} an attempt to make a step from displacement spectra to pair separation spectra (or, simply, separation spectra) was undertaken successfully. A displacement spectrum technique (utilized in the above mentioned publications) operates with displacements of individual tracer from the start point of its trajectory, so that both processes, turbulent motions and large-scale advection, contribute into this spectrum.  To reduce the influence of advection and estimate
the turbulent diffusivity, the pair separation technique \citep[][]{Monin1975, Lesieur1990} can be applied. Here, distances between two tracers at consecutive moments are calculated. Since large-scale advection is expected to effect both tracers equally, it is eliminated. 
The separation spectra for the solar photosphere were reported for the first time by \citet{Lepreti2012}. Nearly the same value $\gamma \approx 1.47-1.49$ was found on scales 10-500 s for all studied magnetic environments: a coronal hole, a quiet sun area, and an active region plage. Note, that for the same data sets, the index $\gamma$, as derived from the displacement spectra,  was different and  increases from the AR plage ($\gamma=1.48$) area to the QS ($\gamma=1.53$) and to the CH ($\gamma=1.67$) \citep[][]{Abramenko2011}. The observed similarity of separation spectra versus individuality of displacement spectra Lepreti and co-authors explained by possible influence of the detailed structure of the velocity field on single tracers dispersal, whereas pair dispersal reflects the diffusivity in the inertial range of turbulence. Therefore, a comparison between the displacement and separation spectra can provide information on the properties of the dispersal mechanism. This kind of comparison is one of the aims of this study. Note, that when comparing the results of \citet{Gianna2014a} with those of \citet{Gianna2014b}, one can conclude that the displacement spectrum  for the intranetwork on scales 100-10000 s with $\gamma = 1.44$ is more shallow than the corresponding separation spectrum with $\gamma = 1.55$, which does not agree with the above mentioned  tendency inferred from the publications of \citet{Abramenko2011} and \citet{Lepreti2012}.  

As we see, the majority of recent studies on the turbulent regime refer to rather small scales: the time intervals below $10^4$ s (approximately 3 hours) and spatial intervals below 1 mega-meter (Mm). Meanwhile, as it was mentioned above, there exists a vital need of the diffusivity properties in a broad range of scales. The seeing-free non-stop data acquired by the Helioseismic and Magnetic Imager (HMI) onboard the Solar Dynamic Observatory (SDO, \citet{Scherrer2012}) offer a good opportunity to extent the scale interval and analyze the magnetic flux dispersal on scales up to  a day and tens of mega-meters. In the present study, this opportunity is used for vast areas of undisturbed photosphere outside active regions. Both displacement and separation spectra on scales 1000-4$\times 10^4$ s (up to 11 hours) were analyzed for a weakest magnetic environment,  a coronal hole, and a typical supergranula pattern.

 \section{Observations and data analysis}
 \subsection{The SDO/HMI data sets}
 
\begin{figure*}
\includegraphics[width=\columnwidth]{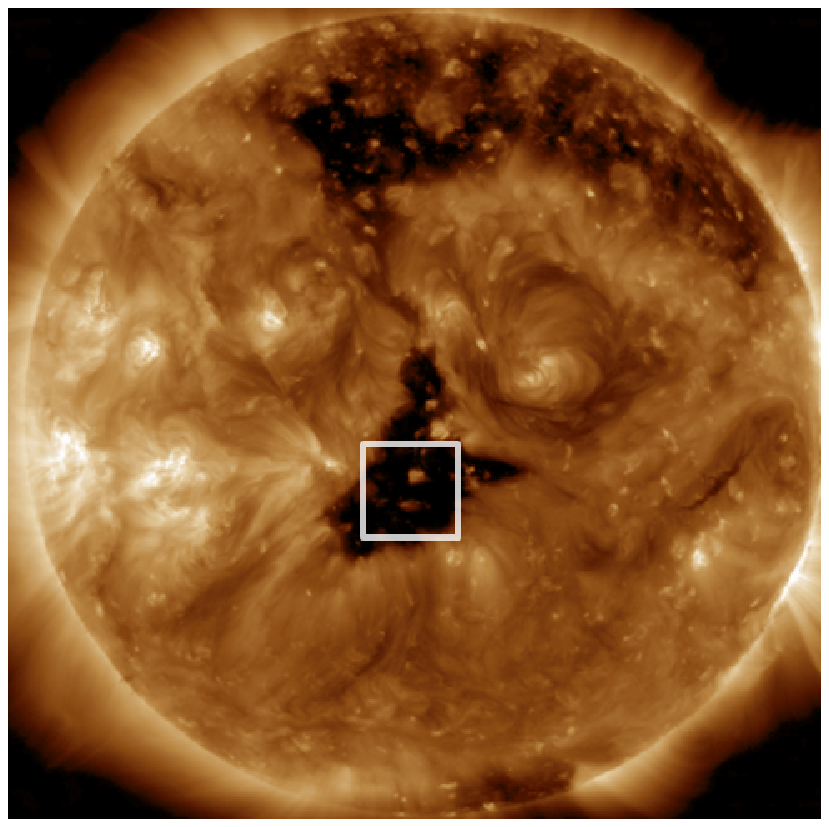}
\includegraphics[width=\columnwidth]{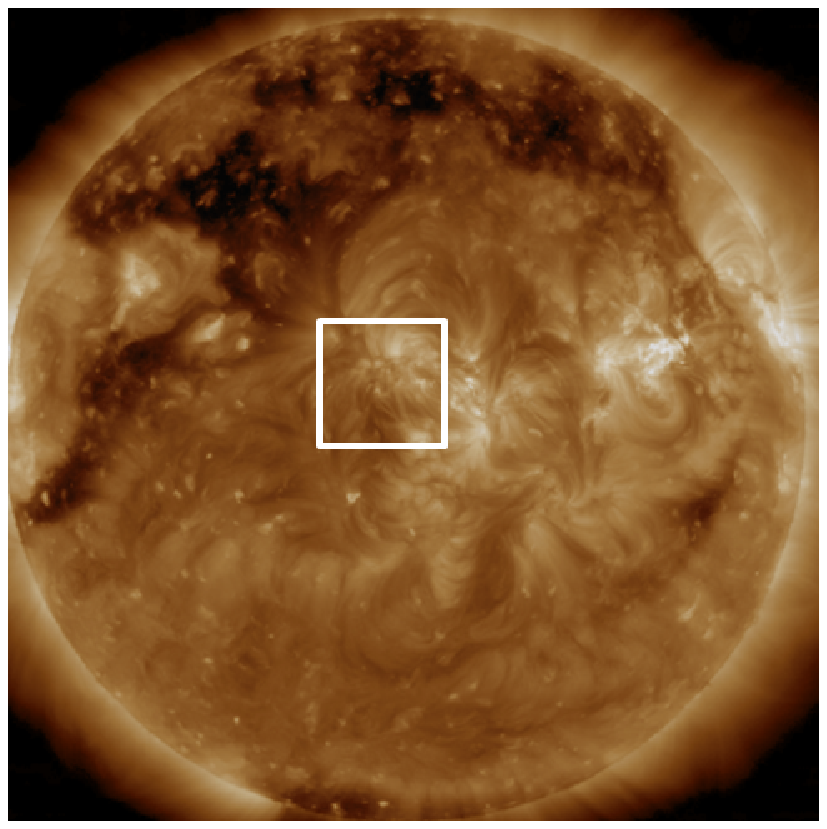}
\caption{\sf Two areas (inside the boxes) analysed in the present study as visible on SDO/AIA 193 A images. Left - data for a coronal hole on the disk center (CH-area) recorded on January 3 at 17:12 UT, 2016; right - data for SG-area recorded on December 1 at 8:36 UT, 2015.     }
\label{fig-1}  
\end{figure*}

Our two data sets consist of magnetogram series obtained with SDO/HMI instrument. The line-of-sight hmi.M-720s magnetograms were taken in the FeI 6173.3 A spectral line with the spatial resolution of 1" (the pixel size of 0.5") and cadence of 12 min \citep{Schou2012} and noise level of about 6  Mx sm$^{-2}$ \citep{Liu2012}.  Two regions of interest were selected (Figure \ref{fig-1}): an area inside a  low-latitude coronal hole (hereinafter CH-area) which crossed the central meridian on January 3, 2016 at approximately 17:12 UT, and an area of decayed active region remains, a typical supergranula pattern (hereinafter SG-area) culminated around December 1, 2015 at 8:36 UT. To avoid the projection effect influence, only one day long intervals from culmination were considered for both cases. For the CH-area, we selected the magnitograms recorded from January 2, 17:12 UT to January 4, 17:00 UT, 2016; for the SG-area, the interval of investigation was November 30, 8:36 UT - December 2, 8:24 UT, 2015. The size of the CH-area was restricted by the boundaries of the coronal hole and consisted 638$\times$636 pixels, or approximately 230$\times$230 Mm. The SG-area covered 758$\times$788 pixels, or 275$\times$286 Mm. The magnetograms taken near the time of culmination are shown in Figure 2. Each data set of 240 magnetograms was carefully aligned using a sub-pixel alignment code based on the fast Fourier transform.

%###########################################################
%#############################################################
\begin{figure}
\includegraphics[width=\columnwidth]{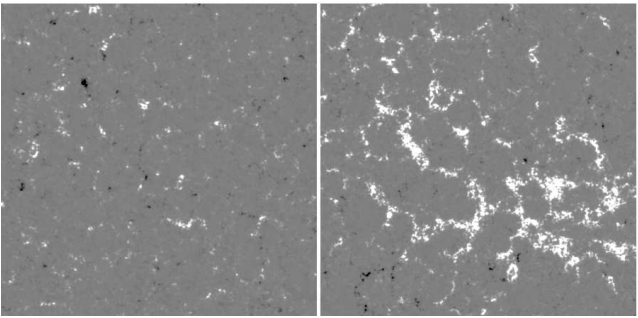}
\caption{\sf Examples of SDO/HMI LOS magnetograms for the CH-area (left) and SG-area (right) recorded at the times depicted in Fig.1. The magnetograms are scaled between -200 (black, negative magnetic polarity) and 200 (white, positive magnetic polarity) Mx sm$^{-2}$. The FOV is 320"$\times$320" (the SG-area is slightly cropped to fit the size of the CH-area). }
\label{fig-2}  
\end{figure}
%#################################################################

\subsection{Detection and tracking of magnetic flux concentrations}

To detect magnetic elements and calculate their trajectories, we applied the modified feature detection and tracking code elaborated by \citet{Abramenko2010, Abramenko2011} for tracking photospheric magnetic bright points. In this study, we used the absolute value of the magnetic field as input. The thresholding technique was applied to obtain a mask of magnetic elements. To count the weakest observed elements and at the same time to mitigate an influence of noise, we choose the threshold of $th=20$ Mx sm$^{-2}$, which corresponds to the triple noise level. A range of sizes of detected elements was selected between 3 and 100 square pixels. 

Each detected element was labeled, the barycenter $(x_c,y_c)$ and equivalent diameter $d$ were calculated, and its counterpart on the consecutive magnetogram (if any) was found. When in the current image, inside the radius of $d/2$ around the $(x_c, y_c)$ pixel, we find a pixel labeled as a center of a magnetic element in the next image, we assign the two found objects to be the same magnetic element visible on two consecutive images. When an element merged with another one, or did not appear on tree consecutive magnetograms, the tracking was terminated. The procedure gives us trajectories of elements. 

As for the coronal hole area, the above procedure results in Set 1 data, i.e., small magnetic elements trajectories inside the CH area (parameters of our data sets are listed in Tables \ref{tab1} and \ref{tab2}).

\begin{table}
\caption{Analyzed data sets}
 \label{tab1}
\begin{tabular}{lrccc}
\hline
Name &  Area  & Size of elements,      & Threshold,          & $N^a$   \\  
     &        & sq. pixels             &  Mx sm$^{-2}$       &         \\
\hline
Set 1 & CH  & 3-100  & 20   & 10591 \\ 
Set 2 & CH  & 20-400 & 130  &   212 \\ 
Set 3 & SG  & 3-100  & 20   & 15635 \\
Set 4 & SG  & 20-400 & 130  & 1602  \\

\hline
\multicolumn{5}{l}{$^a$ Number of tracked elements}\\
\end{tabular}
\end{table}

\begin{table*}
\caption{Calculated parameters}
\label{tab2}
\begin{tabular}{lccccccccc}
\hline
Name &    Lin. range, &  Lin. range$^a$, & $\gamma_d$ & $\gamma_s$ & $K_d(1)^b$ & $K_d(2)^b$ & $K_s(1)^b$ & $K_s(2)^b$   \\  
     &    $10^4$ s    &  $10^3$ km       &            &            &  & &  &    \\ 
\hline
Set 1 &   0.072-3.96 & 0.45-6.05 & 1.333$\pm$0.001 & 1.294$\pm$0.003 & 99  &  378  & 93   & 303  \\
Set 2 &   0.14-1.51  & 0.55-2.21 & 1.258$\pm$0.016 & 1.184$\pm$0.011 & 74  &  136  & 62   & 96   \\
Set 3 &   0.072-3.96 & 0.42-6.32 & 1.364$\pm$0.006 & 1.347$\pm$0.011 & 96  &  416  & 85   & 343  \\
Set 4 &   0.14-1.51  & 0.91-2.56 & 0.880$\pm$0.015 & 0.878$\pm$0.012 & 142 &  108  & 127  & 95   \\
Set 4 &   1.51-3.96  & 2.53-4.80 & 1.249$\pm$0.014 & 1.328$\pm$0.010 & 156 &  196  & 143  & 194  \\

\hline
\multicolumn{3}{l}{$^a$ The spatial linear range for the separation spectra}\\
\multicolumn{6}{l}{$^b$ The diffusion coefficients in km$^2$ s$^{-1}$}\\
\end{tabular}
\end{table*}

To explore dispersion of large magnetic elements, higher values of the threshold and size were selected. Our experience shows that the best choice to detect magnetic elements forming the super-granula boundaries, i.e., network (NW) ensemble, is the threshold of 130 Mx sm$^{-2}$ and the size range of 20-400 square pixels. For the CH area, this procedure gives us the Set 2 data (see Table \ref{tab2} for the calculated parameters). 

Correspondingly, for the SG-area of well-pronounced network pattern (see Fig. \ref{fig-2}), we obtained Set 3 for small elements and Set 4 for large elements. The later represents the majority of the network elements nested on the boundaries of super-granules.  Parameters of data sets are in Tables \ref{tab1}and \ref{tab2}.  

\subsection{Calculation of the displacement and separation spectra}

To analyze the diffusive properties of tracers in a turbulent flow, the Lagrangian approach \citep[e.g.,][]{Monin1975} is usually applied \citep{Abramenko2011, Lepreti2012, Gianna2013, Gianna2014a, Gianna2014b}. Here, the position of a tracer along its trajectory is measured at discrete moments
$t_0, t_1, ... t_i,...t_N$, where $t_0$ is the moment when the tracer was
detected for the first time. Then, the time intervals $\tau_i=t_i - t_0$, from the starting
moment, $t_0$, to the current moment, $t_i$, are calculated for all tracers.
Note that the time moments $t_i$ correspond to times when solar data were
recorded. The next step is to compute the spatial displacements, $(\Delta l)_{i}$, of an
individual $j$-th tracer as a function of time, $\tau_i$. After that, we
calculate the average (over all tracers) displacement for each $\tau_i$ to
produce the average squared displacements (the displacement spectrum):
\begin{equation}
\langle(\Delta l)^{2}(\tau_i)\rangle=\langle|{\bf X}_{j}(0)-{\bf
X}_{j}(\tau_i)|^{2}\rangle,
\label{Delta2}
\end{equation}
where ${\bf X}_{j}(0)=(x_{0},y_{0})$ and ${\bf
X}_{j}(\tau_i)=(x_{\tau_i},y_{\tau_i})$ are coordinates of the $j$-th tracer at
the moment of its first detection and $\tau_i$ seconds later, respectively.
Both processes, advection and turbulent diffusion, contribute into this
spectrum named hereinafter as displacement spectrum.

There are ways to significantly reduce the influence of advection and estimate
the turbulent diffusivity. We utilize for that a widely accepted pair
separation technique \citet{Monin1975, Lepreti2012} keeping in mind that the two-particle dispersion reflects the diffusivity properties arising from the inertial range of turbulence. Here, displacements are
computed as distances between two tracers at consecutive moments.  The pair separation spectrum can be calculated as
\begin{equation}
\langle(\Delta l)^{2}(\tau_i)\rangle=\langle(|{\bf X}_{j}(t_i)-{\bf
X}_{k}(t_i)|  - |l_0^{jk}|)^{2}\rangle,
\label{Pair}
\end{equation}
where $j$ and $k$ denote two tracers, $t_0$ is the first moment when both
tracers are first detected with $l_0^{jk}$ distance between them, and
 $\tau_i = t_i - t_0$. Hereinafter we refer to this kind of spectrum as separation spectrum. 

The power index, $\gamma$, of the spectrum is defined as 
\begin{equation}
\langle(\Delta l)^{2}(\tau)\rangle \sim \tau^{\gamma}
\end{equation}
and is determined as the slope of the spectra over a range of $\tau$. When considering a displacement spectrum, the index will be noted as $\gamma_d$, and, correspondingly, as $\gamma_s$ for a separation spectrum.

If the spectrum that we seek is indeed a power law with index
$\gamma$, and the photospheric plasma is in the turbulence state, then the
dependence of the diffusion coefficient, $K$, from the spatial and time scales
can be derived \citep{Monin1975, Lesieur1990}. Changes of $K$ with scales,
in turn, will help us shed light on the diffusive regime. An expression for the
turbulent diffusion coefficient is
\begin{equation}
K(\tau)=\frac{1}{2D}\frac{d}{d\tau}\langle (\Delta l)^{2}(\tau)\rangle,
\label{Monin_K}
\end{equation}
where $D$ is equal to 2 (3) for diffusion over a surface (volume). Here,
$\langle (\Delta l)^{2}(\tau)\rangle$ is the observed 
spectrum (Eqs. \ref{Delta2}, \ref{Pair}), which can be approximated on a given range of scales as (Abramenko et al. 2011):
\begin{equation}
\langle (\Delta l)^{2}(\tau)\rangle=c\tau^{\gamma},
\label{gen}
\end{equation}
where $c=10^{y_{sect}}$. Values of $\gamma$ and $y_{sect}$ can be derived
from the best linear fit to the spectral data points plotted in a
double-logarithmic plot. Then, accepting that for the diffusion
over the solar surface $D$ equals 2, an expression for the diffusion coefficient was obtained in Abramenko et al. (2011):
\begin{equation}
K(\tau)=\frac{c\gamma}{4}\tau^{\gamma-1}.
\label{Ktau}
\end{equation}
When $\tau$ is excluded from Eqs. \ref{Monin_K} and \ref{gen}, a
relationship between the diffusion coefficient and the spatial scale can be written as (Abramenko et al. 2011):
\begin{equation}
K(\Delta l)=\frac{c\gamma}{4}((\Delta l)^{2}/c)^{(\gamma-1)/\gamma}.
\label{K_l}
\end{equation}
Eqs.\ref{Ktau} - \ref{K_l} show that index $\gamma <1$ leads to an inverse
dependence of the diffusion coefficient on the spatial, $\Delta l$, and
temporal, $\tau$, scales (sub-diffusion). Whereas conditions with $\gamma >1$,
cause $K$ to be directly proportional to scales (super-diffusion).

\section{Results}

\begin{figure}
\includegraphics[width=\columnwidth]{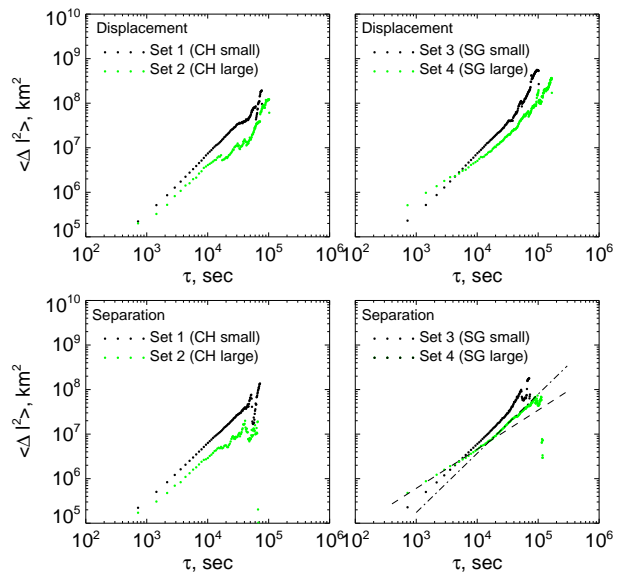}
\caption{\sf Displacement and separation spectra for small magnetic elements (Sets 1 and 3) compared for those for large magnetic elements (Sets 2 and 4). In the right bottom frame, the dashed line shows the best linear fit to the separation spectrum for Set 4 on scales (1440-15100)s with $\gamma = 0.88$, and the dash-dot-dash line shows the best linear fit on scales (15100-39600) s with $\gamma = 1.33$. }
\label{fig-3}  
\end{figure}

\begin{figure}
\includegraphics[width=\columnwidth]{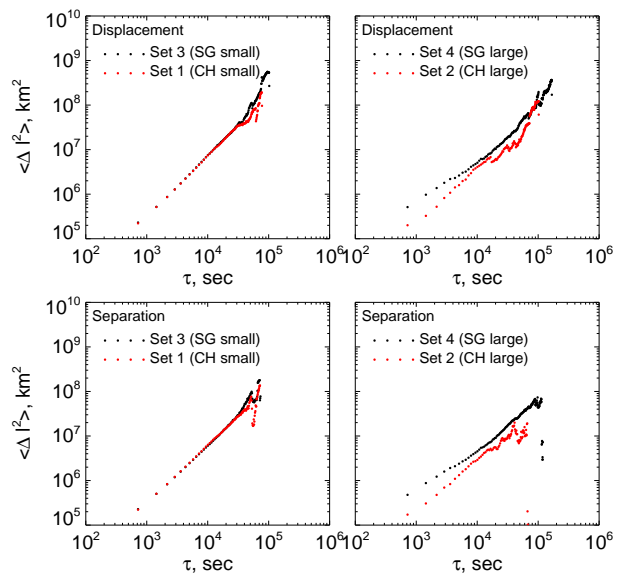}
\caption{\sf  Spectra for the CH area (Sets 1 and 2) compared to the spectra for the SG area (Sets 3 and 4). }
\label{fig-4}  
\end{figure}

\begin{figure}
\includegraphics[width=\columnwidth]{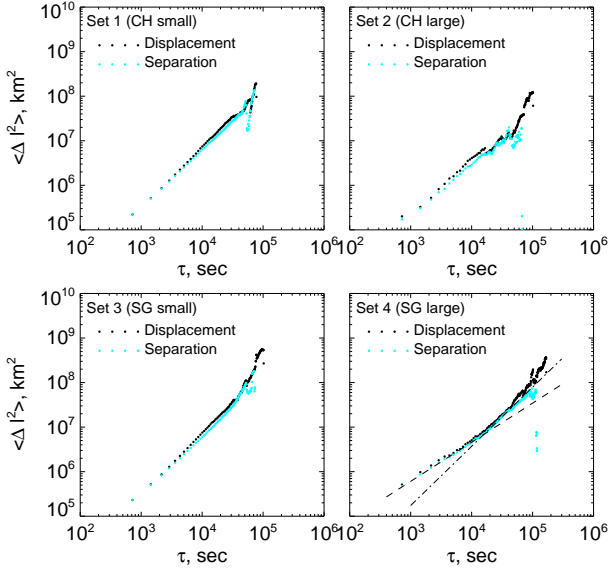}
\caption{\sf The displacement spectra compared to the separation spectra.  Notations are the same as in Fig. \ref{fig-3}. }
\label{fig-5}  
\end{figure}

For each of the four data sets, we calculated two types of spectra: the displacement spectrum (Eq. \ref{Delta2}) and the separation spectrum (Eq. \ref{Pair}). Various combinations to compare them to each other are presented in Figs. \ref{fig-3} - \ref{fig-5}. Hereinafter the parameters for the displacement (separation) spectrum are marked with the subscriber $d$ ($s$).

Does the dispersion of small magnetic elements differ from that of large magnetic elements? To answer, in Figure \ref{fig-3} the spectra for small elements are overplotted by the spectra for large elements. The spectra differ significantly. For the coronal hole area, CH (left panels), the spectra for small elements (Set 1) are above the spectra for large elements (Set 2) on all scales. This implies that the small-scale (mainly intranetwork) magnetic elements in a coronal hole are more mobile than those forming the large-element subset, which supposedly forms the super-granula boundary skeleton. The spectral indices for Set 1 can be defined in a broad range of time scales: (720-39600)~s, which corresponds to 0.45-6.05 Mm for the separation spectrum and 0.46-6.67 Mm for the displacement spectrum, according to Eq. \ref{gen}. The indices are $\gamma_d=1.33 $ and $\gamma_s=1.30$, see Table \ref{tab2}. At the same time, rather low statistics for Set 2 data (only 212 elements) did not allow us to calculate the indicies for the same linear range. Here, values  $\gamma_d=1.26 $ and $\gamma_s=1.18$ were retrieved for the narrow range of (1440-15100) ~s, see Table \ref{tab2}. 

Analyzing the SG-area (right panels in Figure \ref{fig-3}), we found that Set 3 (small elements) reveals a broad linear range of 720-39600 s of super-diffusivity with $\gamma \approx 1.35$, whereas an ensemble of large elements (Set 4) shows two patterns of linearity: on scales 1400-15100 s we observe the sub-diffusion with $\gamma \approx 0.88$ and a decreasing diffusion coefficient (see Table 2), and on scales 15100-39600 s the super-diffusion regime is visible with $\gamma  \approx 1.25-1.33$. The change of the regime occurs on linear scales of approximately 2.5-3 Mm (see Table \ref{tab2} for the linear range). 
Besides, on time scales lower than approximately 5000-8000 s, the large elements are more mobile than the small elements, and the opposite picture is observed on higher time scales. We might conclude that large magnetic elements that compound the super-granular boundaries disperse in a different way than the small magnetic elements.

%##################################################################  
%#################################################################
%##################################################################  
%#################################################################
%##################################################################  
%#################################################################

In Figure \ref{fig-4}, the spectra for the CH area are compared with their counterparts for the SG area. The left panels show that for both displacement and separation spectra, the small elements in the CH area disperse exactly in the same way as they do in the SG area. The difference appears when we consider the dispersion of large elements which form mainly the network skeleton. On all scales, large elements in the CH area being rather scanty (only 212 events) move slower than that in the SG area which are much more numerous (1602 events). On time scales below 15100 s, we observe sub-diffusivity in the SG area and super-diffusivity in the CH area. At the same time, in spite of "accelerated" dispersion (super-diffusivity), the CH large elements (Set 2) display the lower magnitudes of the diffusion coefficient (for example, $K_s $ varies from 62 to 96 km$^2$ s$^{-1}$) as comparing to that for the SG large elements (Set 4,  $K_s $ decreases from 127 to 95 km$^2$ s$^{-1}$, see Table \ref{tab2}).  
   
Figure \ref{fig-5} demonstrates that the displacement spectra are very close to the separation spectra for all data sets (the correspondence is slightly weaker only for the low-statics Set 2). This implies that on time scales below $4\cdot 10^4$ s , or $\sim$11 hours and spatial scales below $\sim$ 6 Mm, large-scale, quasi-regular  patterns of the photospheric horizontal velocity field do not affect the magnetic flux dispersion.  

\begin{figure}
\includegraphics[width=\columnwidth]{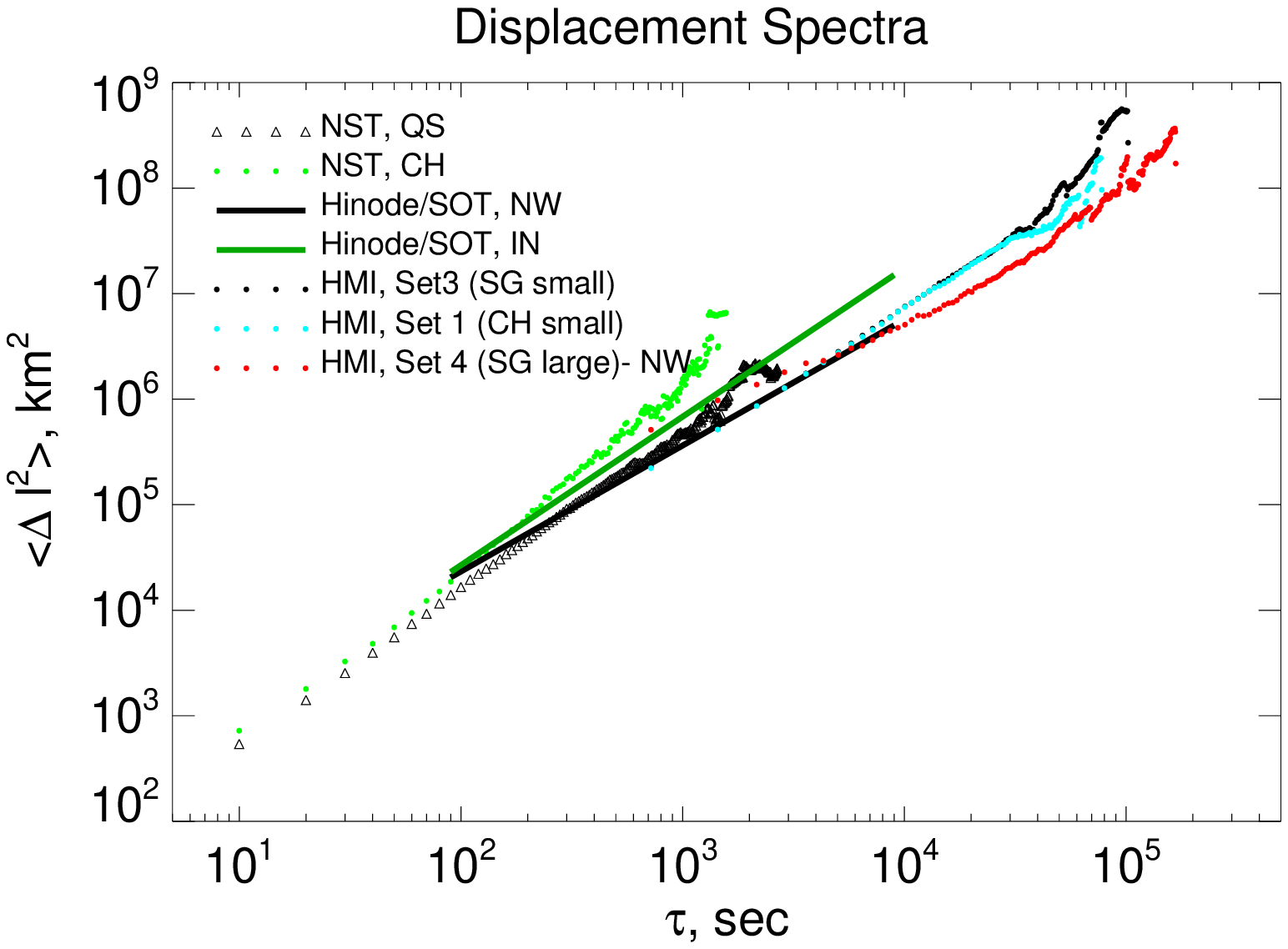}
\caption{\sf Displacement spectra obtained using the data from different instruments. Triangles and green circles denote the NST/BBSO results for quiet sun and coronal hole regions, respectively (from \citet{Abramenko2011}, (reproduced by permission of the AAS). Solid black and green lines schematically represent the result of \citet{Gianna2014a} (reproduced by permission of the AAS) for the network (NW) and intranetwork (IN) ares, respectively. Black and blue dots show the HMI-spectra obtained in the present study for small elements in the SG and CH areas (Sets 1 and 3), respectively, whereas the red dots represent the spectrum from large elements in the SG area (Set 4).      }
\label{fig-6}  
\end{figure}

\begin{figure}
\includegraphics[width=\columnwidth]{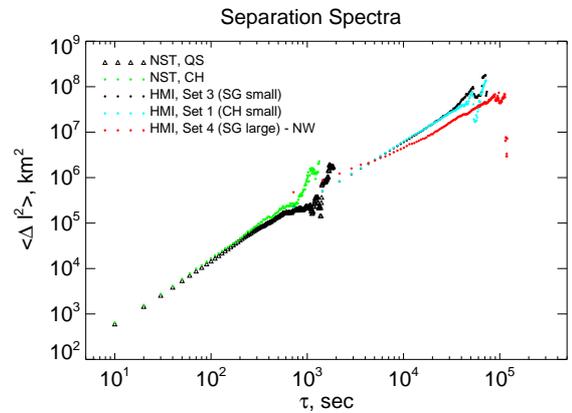}
\caption{\sf Separation spectra for the same data sets as shown in Fig. \ref{fig-6}. NST/BBSO spectra are from the paper by \citet{Lepreti2012} (reproduced by permission of the AAS).   }
\label{fig-7}  
\end{figure}

In Figures \ref{fig-6} and \ref{fig-7} the spectra obtained in this study are overplotted with the previously published data. A general tendency is well pronounced: The spectrum becomes more shallow as the scale increases, i.e., the index $\gamma$ reduces and the regime of well-developed super-diffusivity tends to become closer to the normal diffusion on larger scales. The data reported in \citet{Gianna2014a} are in a good agreements with ours (see Fig. \ref{fig-6}): the transition between the NST- and  HMI-spectra is well covered by the Hinode data. 

\begin{figure}
\includegraphics[width=\columnwidth]{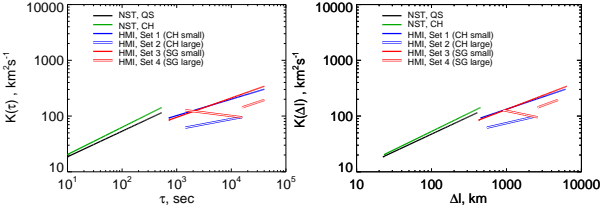}
\caption{\sf  Left panel: diffusion coefficient as a function of temporal scale. Rhight panel: diffusion coefficient as a function of spatial scale.  }
\label{fig-8}  
\end{figure}

The diffusion coefficients as derived from the separation spectra are presented in Figure \ref{fig-8} along with the similar data from the NST-observations reported by \citet{Lepreti2012}. The coefficient increases with scales for all data sets (except Set 4), however, the growth rate becomes slower on larger scales. An abrupt break on scales between the two instruments coverage ($\sim $ 500-700 s and $\sim$ 500 km) might be artificial when the Hinode data are taken into account (see Fig \ref{fig-6}). Anyway, a fair agreement between tiny magnetic bright points (NST data set) and small HMI-magnetic elements (Sets 1 and 3) is noticeable. However, large magnetic elements (Sets 2 and 4, double lines in Fig. \ref{fig-8}) demonstrate the significantly lower values of $K$, which implies the suppressed flux dispersion inside the supergranula boundaries relative to the intergranular zones. 

%##################################################################  
%#################################################################
%##################################################################  
%#################################################################
%##################################################################  
%#################################################################

\section{Conclusions}

 Utilizing the HMI 720 s line-of-sight magnetograms for two regions in the undisturbed photosphere (a coronal hole area and an area of a decayed active region, i.e., a super-granulation pattern), we explored the behavior of the turbulent diffusion coefficient on time scales of approximately 1000-40000~s and spatial scales of approximately 500-6000 km. We analyzed separately the dispersion of small and large magnetic elements. We came to the following inferences.
 
  -  Displacement and separation spectra are very similar to each other for all analyzed data sets, which allows us to suggest that possible influence of large-scale velocity patterns is negligible for the magnetic flux dispersion on scales of interest and, therefore, the inertial range turbulence is explored. 
  
  -  Small magnetic elements in both CH and SG areas disperse in the same way  and they are more mobile than the large ones.  The regime of super-diffusivity is found for them ($\gamma  \approx 1.3 $ and K growths from $\sim$100 to $\sim$ 300 km$^2$ s$^{-1}$). Thus, the hypothesis suggested for the first time in Schrijver et al. (1996) is confirmed by modern observations: large magnetic elements are indeed less mobile than the small ones.  
   
  -  Large magnetic elements in both CH and SG areas disperse slower than the small elements. In the CH area they are scanty and show super-diffusion with $\gamma \approx 1.2$ and $K_s = (62-96)$ km$^2$ s$^{-1}$ on rather narrow scale range of 500-2200 km. Large elements of the SG area demonstrate a band in the spectra and, as a consequence, two ranges of linearity and two diffusivity regimes:  the sub-diffusivity on scales (900-2500) km with $\gamma=0.88$ and $K$ decreasing from $\sim$130 to $\sim$100 km$^2$ s$^{-1}$, and the super-diffusivity on scales (2500-4800) km with $\gamma \approx 1.3$ and $K$ growing from $\sim$140 to $\sim$200 km$^2$ s$^{-1}$. 

The observed here sub-diffusion for large magnetic elements on small scales in the SG-area, on the contrary to super-diffusivity in the CH-area on the same scales, can be interpreted as follows.  We might suggest that widely scattered large elements in the CH hardly form any rigid skeleton of super-granulation, instead they rather freely disperse in the similar regime of super-diffusion as the neighbor small elements do, however with lower coefficients of diffusion. A different situation we observe inside the SG area. Here, the skeleton of network seems to play a role of some constrain factor preventing the "accelerated" dispersion in a super-diffusivity way. Here, magnetic elements are forced to reduce their capability to displace while they walk inside the boundary of SG (on larger scales the super-diffusivity regime is restored). Deep roots and possible inter-connectivity of magnetic flux tubes forming the SG skeleton might be in favor of the observed peculiarities of the flux dispersion. This inference qualitatively agrees with results and conclusions of \citet{Gianna2014a}: the regime of super-diffusivity becomes closer to normal diffusion on scales   $l> 1500$ km and $\tau > 2000$ s creating thus more favorable conditions for accumulation of magnetic flux at the boundaries of super-granules. 

Comparison of our results with the previously published shows that there is a tendency of saturation of the diffusion coefficient on large scales, i.e., the turbulent regime of super-diffusivity gradually ceases so that normal diffusion with a constant value of $K \approx$ 500 km$^2$ s$^{-1}$  might be observed on time scales longer than a day. We presume that only strong and large magnetic elements (capable to survive so long) can be a subject of the expected random walk. However, hardly the used here technique can be applied directly to explore the flux dispersion on such large scales because the basic assumption on the passive nature of the magnetic flux tracers in the turbulent flow might not be applicable on these scales \citep[see, e.g.,][]{Gianna2014a}.

\section*{Acknowledgements}

SDO is a mission for NASA Living With a Star (LWS) program. The SDO/HMI data were provided by the Joint Science Operation Center (JSOC). The study was supported in part by the Russian Foundation for Basic Research projects 16-02-00221 A, 17-02-00049 and 17-52-53203. 

%%%%%%%%%%%%%%%%%%%%%%%%%%%%%%%%%%%%%%%%%%%%%%%%%%



%\bibliographystyle{mnras}


%%%%%%%%%%%%%%%%%%%%%%%%%%%%%%%%%%%%%%%%%%%%%%%%%%

% Don't change these lines
\bsp	% typesetting comment
\label{lastpage}
\end{document}